\documentclass[twocolumn]{aastex62}

\pdfoutput=1

\begin{document}

\title{THE GALAXY'S GAS CONTENT REGULATED BY THE DARK MATTER HALO MASS RESULTS IN A SUPER-LINEAR M$_{\rm BH}$--M$_{\star}$ RELATION}

\correspondingauthor{Ivan Delvecchio}
\email{ivan.delvecchio@cea.fr}

\author{I. Delvecchio}\altaffiliation{Marie Curie Fellow}
\affil{CEA, IRFU, DAp, AIM, Universit\'e Paris-Saclay, Universit\'e Paris Diderot, Sorbonne Paris Cit\'e, CNRS, F-91191 Gif-sur-Yvette, France}
\affil{INAF - Osservatorio di Astrofisica e Scienza dello Spazio, via Gobetti 93/3, I-40129,Bologna, Italy}
\author{E. Daddi}
\affil{CEA, IRFU, DAp, AIM, Universit\'e Paris-Saclay, Universit\'e Paris Diderot, Sorbonne Paris Cit\'e, CNRS, F-91191 Gif-sur-Yvette, France}
\author{F. Shankar}
\affil{Department of Physics and Astronomy, University of Southampton, Highfield SO17 1BJ, UK}
\author{J. R. Mullaney}
\affil{Department of Physics and Astronomy, The University of Sheffield, Hounsfield Road, Sheffield S3 7RH, UK}
\author{G. Zamorani}
\affil{INAF - Osservatorio di Astrofisica e Scienza dello Spazio, via Gobetti 93/3, I-40129,Bologna, Italy}
\author{J. Aird}
\affil{Department of Physics \& Astronomy, University of Leicester, University Road, Leicester LE1 7RJ, UK}
\author{E. Bernhard}
\affil{Department of Physics and Astronomy, The University of Sheffield, Hounsfield Road, Sheffield S3 7RH, UK}
\author{A. Cimatti}
\affil{University of Bologna, Department of Physics and Astronomy (DIFA), Via Gobetti 93/2, I-40129, Bologna, Italy}
\affil{INAF - Osservatorio Astrofisico di Arcetri, Largo E. Fermi 5, I-50125, Firenze, Italy}
\author{D. Elbaz}
\affil{CEA, IRFU, DAp, AIM, Universit\'e Paris-Saclay, Universit\'e Paris Diderot, Sorbonne Paris Cit\'e, CNRS, F-91191 Gif-sur-Yvette, France}
\author{M. Giavalisco}
\affil{Department of Astronomy, University of Massachusetts Amherst, 710 North Pleasant Street, Amherst, MA 01003-9305, USA}
\author{L. P. Grimmett}
\affil{Department of Physics and Astronomy, The University of Sheffield, Hounsfield Road, Sheffield S3 7RH, UK}

\begin{abstract}

Supermassive black holes (SMBHs) are tightly correlated with their hosts but the origin of such connection remains elusive. To explore the cosmic build-up of this scaling relation, we present an empirically-motivated model that tracks galaxy and SMBH growth down to z=0. Starting from a random mass seed distribution at z=10, we assume that each galaxy evolves on the star-forming ``main sequence'' (MS) and each BH follows the recently-derived stellar mass (M$_{\star}$) dependent ratio between BH accretion rate and star formation rate, going as BHAR/SFR~$\propto$~M$_{\star}^{0.73[+0.22,-0.29]}$. Our simple recipe naturally describes the BH-galaxy build-up in two stages. At first, the SMBH lags behind the host that evolves along the MS. Later, as the galaxy grows in M$_{\star}$, our M$_{\star}$--dependent BHAR/SFR induces a super-linear BH growth, as M$_{\rm~BH}$~$\propto$~M$_{\star}^{1.7}$. According to this formalism, smaller BH seeds increase their relative mass faster and earlier than bigger BH seeds, at fixed M$_{\star}$, thus setting along a gradually tighter M$_{\rm~BH}$--M$_{\star}$ locus towards higher M$_{\star}$. Assuming reasonable values of the radiative efficiency $\epsilon \sim$0.1, our empirical trend agrees with both high-redshift model predictions and intrinsic M$_{\rm~BH}$--M$_{\star}$ relations of local BHs. We speculate that the observed non-linear BH-galaxy build-up is reflected in a twofold behavior with dark matter halo mass (M$_{\rm DM}$), displaying a clear turnover at M$_{\rm DM}\sim$2$\times$10$^{12}$~M$_{\odot}$. While Supernovae-driven feedback suppresses BH growth in smaller halos (BHAR/SFR~$\propto$~M$_{\rm DM}^{1.6}$), above the M$_{\rm DM}$ threshold cold gas inflows possibly fuel both BH accretion and star formation in a similar fashion (BHAR/SFR~$\propto$~M$_{\rm DM}^{0.3}$). 

\end{abstract}


\section{Introduction} \label{intro}

How supermassive black holes (SMBHs) formed and evolved with cosmic time is one of the most debated issues in modern Astrophysics. One of the best known evidence supporting co-evolution between SMBHs and their host galaxies is the observed relationship at z$\sim$0 between SMBH mass (M$_{\rm BH}$) and several properties of galaxy bulges: stellar velocity dispersion ($\sigma_{*}$), stellar bulge mass (M$_{\rm bulge}$), dark matter halo mass (M$_{\rm DM}$) (e.g. \citealt{Magorrian1998}; \citealt{Gebhardt+2000}; \citealt{Ferrarese+2000}; \citealt{Haring+2004}; \citealt{Gultekin+2009}). 

Such tight (scatter $\sim$0.3~dex) correlation is currently interpreted as the outcome of a long-term balance between feeding and feedback processes occurring in galaxy bulges and their central BHs (see a comprehensive review by \citealt{Kormendy+2013}). 

Nevertheless, still unclear is whether the local M$_{\rm BH}$--M$_{\rm bulge}$ relation observed for classical galaxy bulges (M$_{\rm BH}$/M$_{\rm bulge}<$1/200, \citealt{Kormendy+2013}) evolves with redshift. Several studies targeting high-redshift quasars found BHs as massive as 10$^{9}$~M$_{\odot}$ at z$>$6, when the Universe was less than 1~Gyr old (\citealt{Mortlock+2011}; \citealt{Wu+2015}; \citealt{Banados+2018}; \citealt{Vito+2019}). This suggests the presence of high-redshift BHs that are \textit{overmassive} (M$_{\rm BH}$/M$_{\star} >$1/100) relative to local scaling relations, as found in local giant ellipticals \citep{Lupi+2019}. Nevertheless, their M$_{\rm BH}$ measurements might be biased, since they rely on gas dynamical estimates on kpc scales, which might not hold within the BH sphere of influence. An alternative scenario is that the galaxy stellar/halo mass primarily regulates the amount of cold gas available for triggering and sustaining the central SMBH growth (see \citealt{Volonteri2010} for a review). Investigating the relationship between BH accretion rate (BHAR) and star formation rate (SFR) is crucial to shed light on the connection between both phenomena at various epochs.

A pioneering study of \citet{Mullaney+2012} first proposed the idea that SMBH and galaxy growth are synchronised at all times at a universal BHAR/SFR$\sim$10$^{-3}$. Recently, a number of empirical evidence argued that the BHAR/SFR ratio increases with M$_{\star}$ (\citealt{Rodighiero+2015}; \citealt{Yang+2018}; \citealt{Aird+2019}). This was independently corroborated in \citeauthor{Delvecchio+2019} (\citeyear{Delvecchio+2019}, D19 hereafter), via modeling the observed AGN X-ray luminosity function (XLF, \citealt{Aird+2015}). In this letter, we explore the implications on the cosmic SMBH growth resulting from a M$_{\star}$--dependent BHAR/SFR trend. Particularly, assuming a seed distribution for both M$_{\rm BH}$ and galaxy M$_{\star}$ starting at very high redshift (z=10), we let it evolve following the above trend. Finally, we compare their final mass build-up with observed scaling relations at z=0 and state-of-the-art cosmological simulations at higher redshifts.

Throughout this letter, we adopt a \citet{Chabrier2003} initial mass function (IMF) and a flat cosmology with $\Omega_{\rm m}$=0.30, $\Omega _{\Lambda}$=0.70 and H$_{0}$=70~km~s$^{-1}$~Mpc$^{-1}$.

\section{Our empirically-motivated toy model} \label{method}

\begin{figure}
     \includegraphics[width=\linewidth]{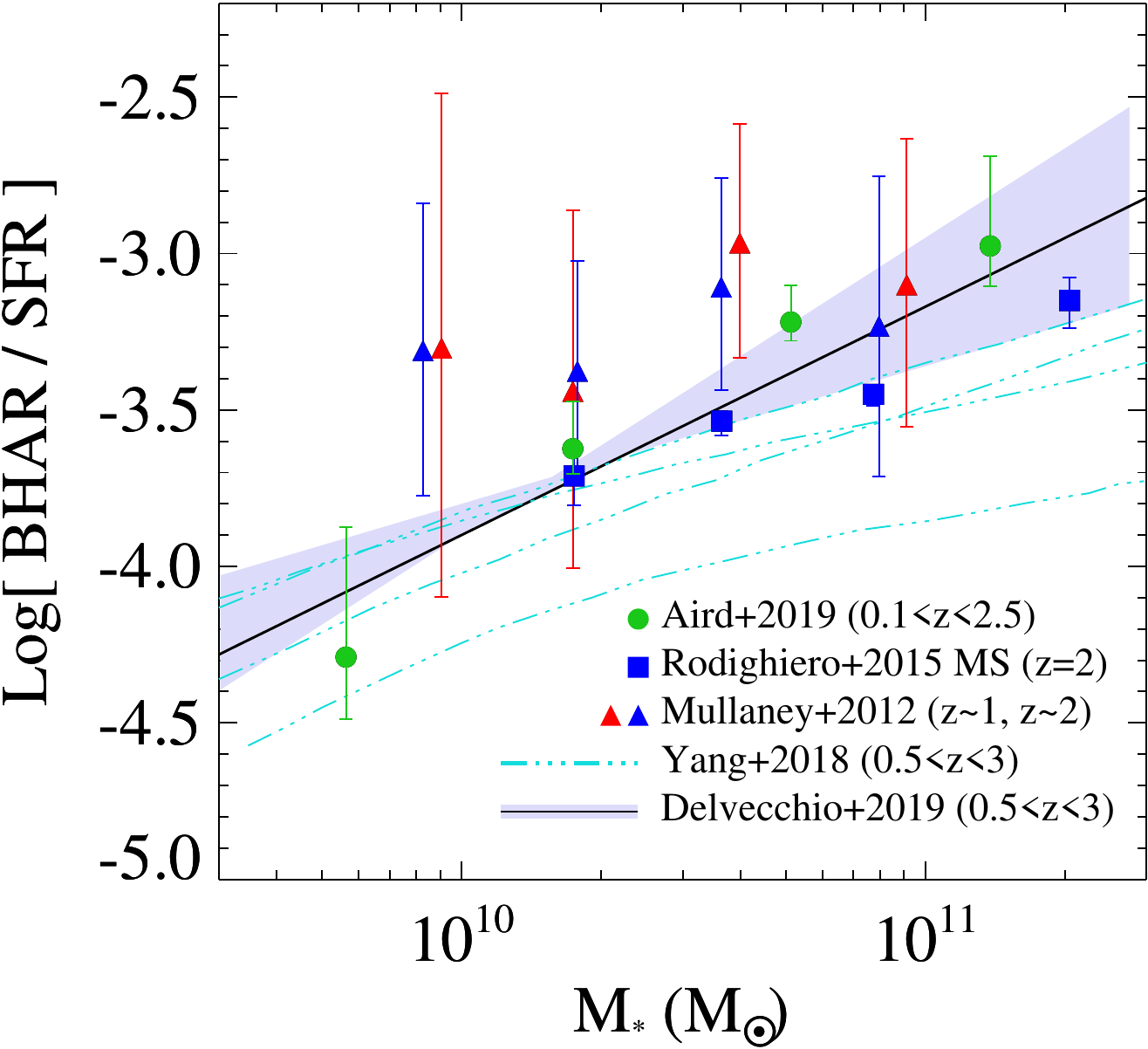}
 \caption{\small Compilation of various mean BHAR/SFR trends with M$_{\star}$ proposed in the literature. Empirical relations for star-forming galaxies are taken from \citeauthor{Mullaney+2012} (\citeyear{Mullaney+2012}, z$\sim$1 and z$\sim$2, red and blue triangles, respectively), \citeauthor{Rodighiero+2015} (\citeyear{Rodighiero+2015}, z$\sim$2, blue squares) and \citeauthor{Yang+2018} (\citeyear{Yang+2018}, dot-dashed lines, increasing with redshift over 0.5$<$z$<$3). Datapoints from \citet{Aird+2019} (green circles) are averaged over the BHAR/SFR distributions across 0.1$<$z$<$2.5. Our recent trend (D19, black solid line) was obtained by reproducing the observed XLF of AGN at 0.5$<$z$<$3, yielding BHAR/SFR~$\propto$~M$_{\star}^{0.73[+0.22,-0.29]}$ at $\pm$1$\sigma$ confidence level (grey shaded area). 
 }
   \label{fig:bhsf}
\end{figure}

D19 successfully reproduced the observed AGN XLF \citep{Aird+2015} since z$\sim$3, disentangling the relative contribution of main-sequence (MS) and starburst (SB) galaxies. The XLF was modeled as the convolution between the galaxy M$_{\star}$ function and a set of ``specific BHAR'' (s-BHAR = BHAR/M$_{\star}$~$\propto$~L$_{\rm X}$/M$_{\star}$, see \citealt{Aird+2012}) distributions, that were normalised to match a number of empirical BHAR/SFR trends. From the derived XLF, we directly constrained the typical BHAR/SFR ratio to scale positively with M$_{\star}$, as BHAR/SFR~$\propto$~M$_{\star}^{0.73[+0.22,-0.29]}$, and roughly independent of redshift at 0.5$<$z$<$3 (e.g. \citealt{Aird+2019}). While extrapolating this BHAR/SFR trend at z$>$3 might suffer from uncertainties, this finding suggests that SMBHs and their hosts do not grow in lockstep over cosmic time.
Fig.~\ref{fig:bhsf} displays our BHAR/SFR trend with M$_{\star}$ (black solid line), and the corresponding $\pm$1$\sigma$ scatter (grey shaded area). For comparison, in Fig.~\ref{fig:bhsf} we report other data and trends from the literature (\citealt{Mullaney+2012}, \citealt{Rodighiero+2015}; \citealt{Yang+2018}; \citealt{Aird+2019}) at various redshifts. In particular, \citet{Yang+2018} argue for a flatter BHAR/SFR trend with M$_{\star}$, and slightly increasing with redshift. However, we stress that the redshift dependence is, at least partly, a consequence of the M$_{\star}$-independent MS relation assumed by the authors (from \citealt{Behroozi+2013}). The absence of a bending towards high M$_{\star}$ leads to slightly higher SFR, therefore lower BHAR/SFR relation, especially at low redshift where the flattening is stronger (e.g. \citealt{Schreiber+2015}). Therefore, under the assumption of a bending MS, the above studies are all consistent with a redshift-invariant BHAR/SFR ratio.

\subsection{Initial mass seed distributions}  \label{seed}

In order not to bias ourselves to any prior seed distribution, we start with a uniform sample of one thousand seeds formed at z=10, with masses M$_{\rm BH}$=10$^{2-6}$~M$_{\odot}$ and M$_{\star}$=10$^{6-10}$~M$_{\odot}$. Such input grid spans a wide range of M$_{\rm BH}$/M$_{\star}$, covering the mass range predicted by the main BH seed formation channels \citep{Begelman+1978}: (i) PopIII stars remnants ($\sim$10$^2$~M$_{\odot}$); (ii) stellar dynamical collapse ($\sim$10$^3$~M$_{\odot}$); (iii) gas dynamical collapse ($\sim$10$^{5-6}$~M$_{\odot}$). Taking a higher (lower) initial redshift would simply yield slightly smaller (larger) M$_{\star}$ at z=0. Independently of this, their M$_{\rm BH}$ estimates would scale accordingly (based on the BHAR/SFR trend with M$_{\star}$), keeping our final results and conclusions unchanged.

\subsection{Setting galaxy M$_{\star}$ growth}  \label{seed_galaxy}

For each galaxy M$_{\star}$ and redshift, we assign the corresponding SFR by following the MS relation of \citet{Schreiber+2015}, re-scaled to a \citet{Chabrier2003} IMF. The MS scatter was propagated on the derived SFR by following a log-normal distribution with 1$\sigma$ dispersion of 0.3~dex \citep{Schreiber+2015}. The cumulative M$_{\star}$ is simply calculated as the time-integral of the SFR. We acknowledge that a more detailed treatment of the M$_{\star}$ build-up would require a correction for stellar mass losses \citep{Leitner+2011}, which would slightly lower our integrated M$_{\star}$, and consequently our M$_{\rm BH}$, without affecting the overall trend. In fact, our main goal is to track the cosmic assembly of the M$_{\rm BH}$--M$_{\star}$ \textit{slope and normalisation} at various epochs, not to match the observed galaxy M$_{\star}$ distribution at each redshift.

\subsection{Setting BH growth}  \label{seed_bh}

Each BH seed is assumed to gain mass via gas accretion \citep{Soltan1982} at a fixed radiative efficiency $\epsilon$=0.1 (e.g. \citealt{Marconi+2004}). Because of a M$_{\star}$--dependent BHAR/SFR ratio, we translate the corresponding SFR into a \textit{long-term average} BHAR, at each M$_{\star}$. Based on this formalism, we determine the cumulative accreted SMBH mass since z=10 as:
\begin{equation}
\rm M_{\rm BH}|_{z=0} = \int_{z=10}^{z=0} BHAR|_{M_{\star}(z')} \cdot \frac{dt'}{dz'} ~ dz' ~~ + ~~ M_{\rm BH}|_{z=10}
\label{eq:bhgrowth}
\end{equation}
The corresponding Eddington ratio $\lambda_{\rm EDD}$, at each M$_{\star}$ (and redshift), is calculated as:
\begin{equation}
\rm \lambda_{\rm EDD}|_{M_{\star}(z')} = 4.41\cdot10^8 \cdot \frac{\epsilon}{1-\epsilon} \cdot \frac{BHAR|_{M_{\star}(z')}}{M_{\rm BH}/M_{\odot}}
\label{eq:eddratio}
\end{equation}
We iterate the above procedure down to z=0 with a redshift step of 0.1.

\begin{figure*}
\centering
\includegraphics[width=5.7in]{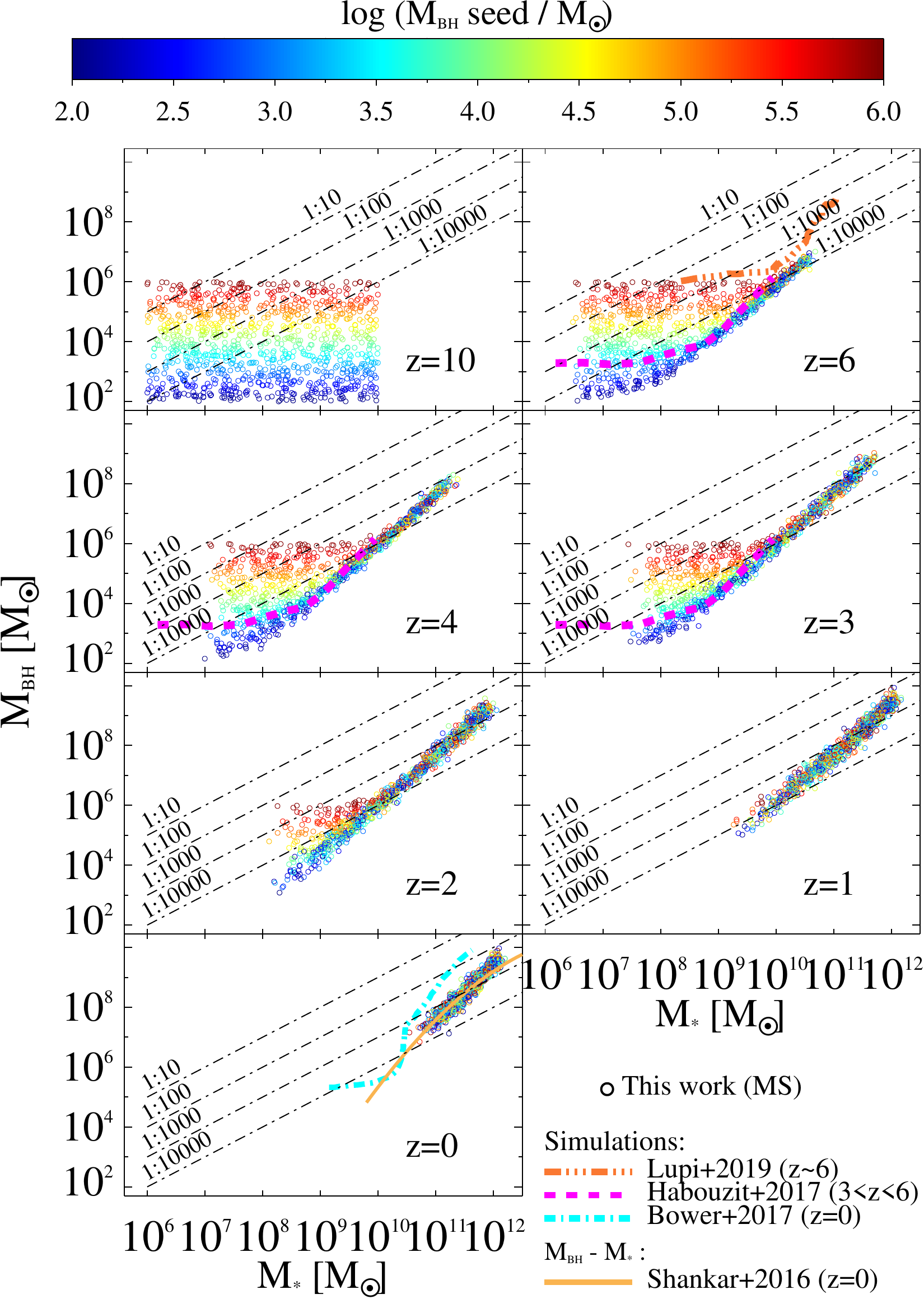}
 \caption{\small Cosmic build-up of SMBH and galaxy mass implied by our M$_{\star}$--dependent BHAR/SFR trend (open circles). We assume uniform M$_{\rm BH}$ and M$_{\star}$ seed distributions at z$_{\rm f}$=10, spanning the ranges 10$^{2}<$M$_{\rm BH}<$10$^{6}$ and 10$^{6}<$M$_{\star}<$10$^{10}$~M$_{\odot}$, respectively (top-left panel). The colorbar indicates the seed M$_{\rm BH}$ distribution at z$_{\rm f}$. Dot-dashed lines mark various M$_{\rm BH}$/M$_{\star}$ ratios. We track the evolution of M$_{\star}$ and M$_{\rm BH}$ for MS galaxies (circles), incorporating the scatter of the MS relation and the uncertainties on the assumed BHAR/SFR trend. For comparison, we show predictions of a SB galaxy at z$\sim$6 (\citealt{Lupi+2019}, orange dot-dashed line), and of normal star-forming galaxies (\citealt{Habouzit+2017}, magenta dotted line at 3$<$z$<$6; \citealt{Bower+2017}, cyan dot-dashed line at z=0). We also show the proposed de-biased M$_{\rm BH}$--M$_{\star}$ relation at z=0 (\citealt{Shankar+2016}, yellow solid line). The absence of galaxies with M$_{\star}<$10$^{10}$~M$_{\odot}$ at z=0 is simply attributable to our limited M$_{\star}$ grid at z=10.
 }
   \label{fig:sample}
\end{figure*}

\begin{figure}
     \includegraphics[width=\linewidth]{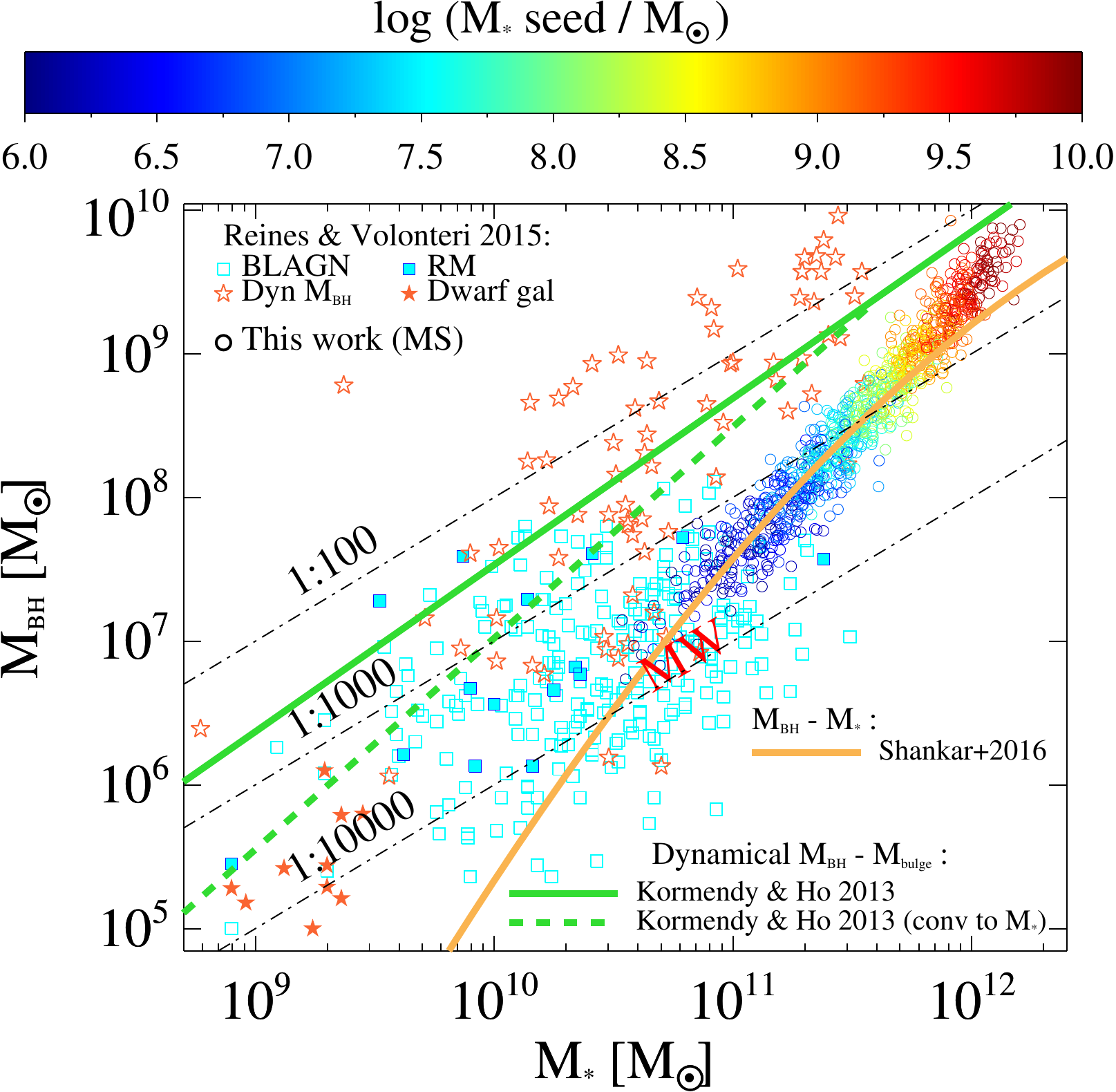}
 \caption{\small Final M$_{\rm BH}$--M$_{\star}$ relation at z=0 determined by our M$_{\star}$--dependent BHAR/SFR. The colorbar indicates the seed M$_{\star}$ distribution at z$_{\rm f}$. For comparison, we show the local relation from \citet{Kormendy+2013}, both with M$_{\rm bulge}$ (green solid line) and converted to total M$_{\star}$ (green dashed line). Our MS wedge agrees remarkably well with the proposed de-biased M$_{\rm BH}$--M$_{\star}$ trend (\citealt{Shankar+2016}, yellow solid line) and with the virial M$_{\rm BH}$ estimates for BLAGN (empty squares, \citealt{Reines+2015}). For completeness, we report M$_{\rm BH}$--M$_{\star}$ estimates collected from \citet{Reines+2015} for reverberation mapping AGN (RM, filled squares), dynamical M$_{\rm BH}$ measurements (empty stars) and dwarf galaxies (filled stars). The tag ``MW'' marks the mass measurements for the Milky Way. }
   \label{fig:z0}
\end{figure}

\section{Results} \label{results}

Exploring the cosmic build-up induced by a M$_{\star}$--dependent BHAR/SFR ratio is paramount for understanding whether the galaxy (halo) governs SMBH growth by setting the available amount of gas for fueling SF and BH accretion, or instead if early AGN feedback controls the amount of cold star-forming gas that fuels galaxy growth \citep{Volonteri2010}. 

Fig.~\ref{fig:sample} displays the evolution of one thousand M$_{\star}$ and M$_{\rm BH}$ seeds (circles) since z$_{\rm f}$=10, resulting from our empirical M$_{\star}$--dependent BHAR/SFR (D19). The colorbar indicates the seed M$_{\rm BH}$ distribution. For convenience, dot-dashed lines mark various M$_{\rm BH}$/M$_{\star}$ ratios. We track the evolution of M$_{\star}$ and M$_{\rm BH}$ while propagating, at each time step, the dispersion of the MS relation and the uncertainty on the BHAR/SFR trend. 

The positive BHAR/SFR relation with M$_{\star}$ suggests that SMBH accretion and star formation do \textit{not} proceed in lockstep at all cosmic epochs, whereas their build-up comes in two stages (Fig.~\ref{fig:sample}). 

Because of our M$_{\star}$--dependent BHAR/SFR trend, in small M$_{\star}$ galaxies the BHAR is quite low relative to the SFR, therefore the SMBH lags behind the galaxy. As the galaxy steadily grows in M$_{\star}$, the BHAR is progressively enhanced relative to the SFR. In this regime, the BH grows super-linearly as M$_{\rm BH}$~$\propto$~M$_{\star}^{1.7}$, setting along a gradually tighter M$_{\rm BH}$--M$_{\star}$ locus towards higher M$_{\star}$. This is because two BHs with different M$_{\rm BH}$ but same galaxy M$_{\star}$ have the same BHAR. Hence, while the \textit{absolute} BH mass gained per unit time is the same, a smaller BH seed will increase its \textit{relative} mass by a much larger factor than a bigger BH seed.

Therefore, the seed M$_{\star}$ is the key quantity that \textit{attracts} all seeds towards a super-linear slope, above a critical M$_{\star}$ value (Fig.~\ref{fig:sample}). Instead, the seed M$_{\rm BH}$ is the parameter that sets the corresponding M$_{\star}$ threshold, increasing with seed M$_{\rm BH}$, above which any prior BH seed dependence is lost. 

For comparison, Fig.~\ref{fig:sample} reports predictions from cosmological simulations of a SB galaxy at z$\sim$6 (\citealt{Lupi+2019}, orange dot-dashed line), and of normal star-forming galaxies (\citealt{Habouzit+2017}, magenta dotted line at 3$<$z$<$6; \citealt{Bower+2017}, cyan dot-dashed line at z=0). \citet{Lupi+2019} used a cosmological simulation for studying the evolution of a quasar host-galaxy at z$\sim$6--7, with seed M$_{\rm BH}$=10$^{6}$~M$_{\odot}$ at z$\gtrsim$10. Instead, the model of \citet{Habouzit+2017} explores the early growth (3$<$z$<$8) of lighter BH seeds, with M$_{\rm BH}$=10$^{2-3}$~M$_{\odot}$. Finally, \citet{Bower+2017} report the results from the EAGLE cosmological simulation \citep{Schaye+2015}.
All these models support strong Supernovae-driven winds in the early phases of galaxy growth, that evacuate the gas around the central SMBH \citep{Dubois+2014}, halting both BH and galaxy bulge growth. However, the rest of the galaxy keeps growing on the MS until it reaches a critical M$_{\star}$, increasing with seed M$_{\rm BH}$. At this stage, the galaxy potential well is deep enough to retain the ejected gas, and to drive it more effectively towards the center. The BH grows super-linearly with M$_{\star}$ matching the proposed de-biased M$_{\rm BH}$--M$_{\star}$ relation (\citealt{Shankar+2016}, yellow solid line, see Fig.~\ref{fig:z0}). This predicted scenario is qualitatively consistent with our empirical toy model predictions, as a natural consequence of a M$_{\star}$--dependent BHAR/SFR ratio. A noteworthy difference is that the above cosmological simulations do not directly link the BHAR to the host's properties, while our toy model assumes that BHAR depends exclusively on the galaxy's star-forming content.

In Fig.~\ref{fig:z0} we further test our empirical predictions at z=0 against the local relation by \citet{Kormendy+2013}, both with M$_{\rm bulge}$ (green solid line) and converted to total M$_{\star}$ (green dashed line), by applying a M$_{\star}$--dependent bulge-to-total (B/T) correction for local MS galaxies \citep{Dimauro+2018}. Our MS wedge agrees remarkably well with the M$_{\rm BH}$--M$_{\star}$ of \citeauthor{Shankar+2016} (\citeyear{Shankar+2016}, yellow solid line), suggesting that our long-term averaged BHAR/SFR trend is able to recover the intrinsic M$_{\rm BH}$--M$_{\star}$ relation of MS galaxies. In addition, our trend fits very well the BLAGN sample of \citet{Reines+2015}, who exploited 262 single-epoch M$_{\rm BH}$ estimates down to $\approx$10$^{5}$~M$_{\odot}$, for which they computed the total galaxy M$_{\star}$. This low-mass AGN sample contains moderate luminosity AGN (10$^{41.5}<$L$_{\rm AGN}<$10$^{44.4}$~erg~s$^{-1}$), hence significantly more common than previous quasars samples. The authors found a 10$\times$ lower normalisation than that inferred for dynamical M$_{\rm BH}$--M$_{\rm bulge}$ trends of inactive elliptical galaxies \citep{Kormendy+2013}, that we interpret in the next Section.

\section{Discussion and Summary} \label{discussion}

Our findings corroborate the idea that galaxies and their SMBHs do not grow in lockstep at all times. Cosmological simulations predict that the SMBH first starves until the galaxy reaches a critical M$_{\star}\sim$10$^{9-10}$~M$_{\odot}$ (\citealt{Bower+2017}; \citealt{Habouzit+2017}; \citealt{Lupi+2019}), corresponding to M$_{\rm DM}\sim$10$^{11-12}$~M$_{\odot}$ (e.g. \citealt{McAlpine+2018}). Later, they predict a super-linear BH growth towards M$_{\rm BH}$/M$_{\star} \gtrsim$10$^{-3}$ at M$_{\star} \gtrsim$10$^{11}$~M$_{\odot}$, in qualitative agreement with our findings.

\begin{figure}
     \includegraphics[width=\linewidth]{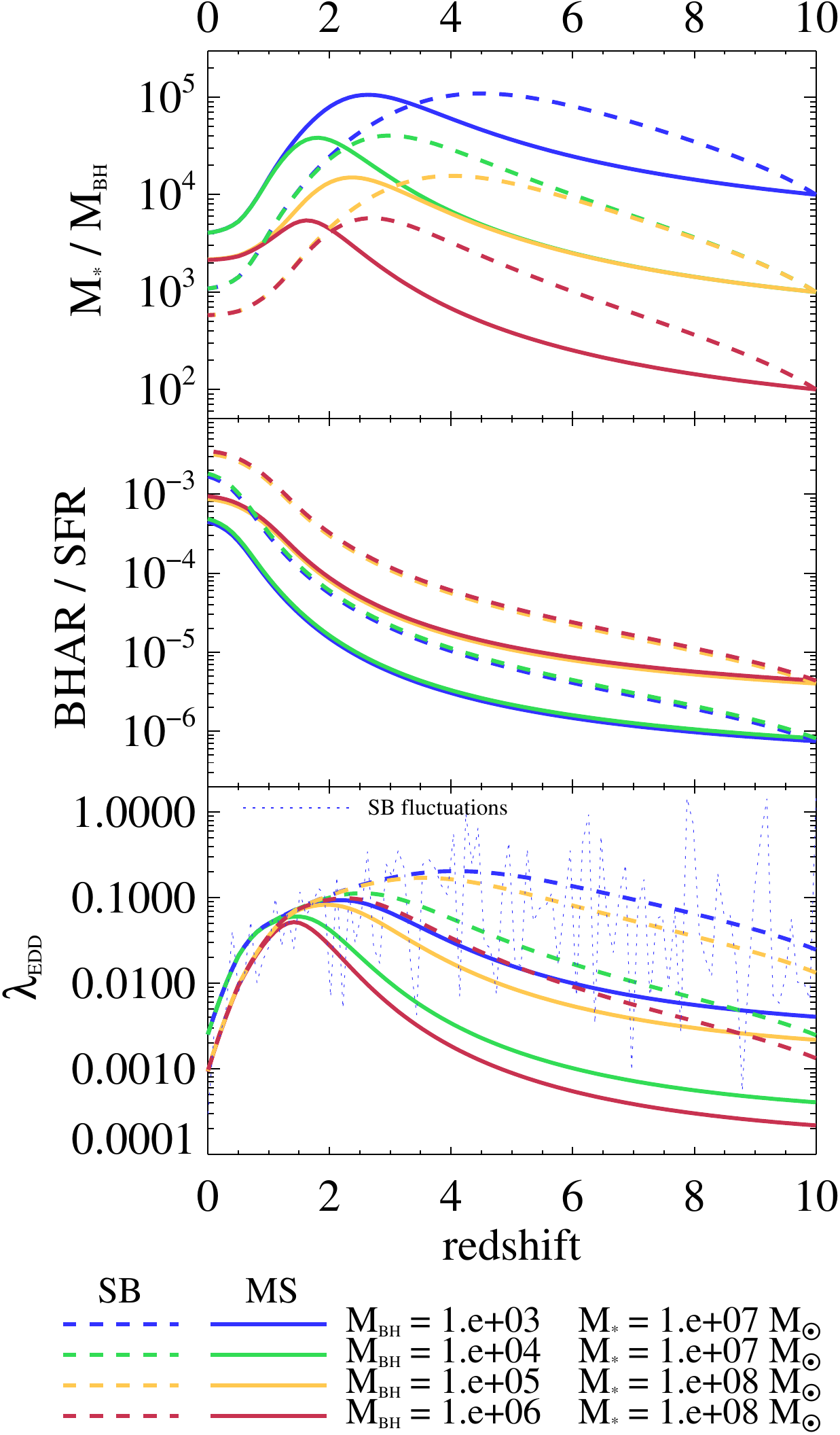}
 \caption{\small Redshift evolution of various SMBH- and galaxy-related parameters for four different seed masses (colored lines). The trends for MS and SB galaxies are highlighted with solid and dashed lines, respectively. Top panel: M$_{\rm BH}$/M$_{\star}$ ratio. Middle panel: BHAR/SFR ratio. Bottom panel: Eddington ratio ($\lambda_{\rm EDD}$). The blue dotted line marks the typical fluctuations in $\lambda_{\rm EDD}$ of a SB galaxy with seed (M$_{\rm BH}$,M$_{\star}$) = (10$^{3}$,10$^{7}$)~M$_{\odot}$ (blue dashed line) when propagating the dispersion of the MS relation and the scatter of the assumed BHAR/SFR trend.
 }
   \label{fig:evo}
\end{figure}

The two-fold trend predicted also by our toy model can be better visualised in Fig.~\ref{fig:evo}, which displays the cosmic evolution of M$_{\rm BH}$/M$_{\star}$ (top panel), BHAR/SFR (middle panel) and $\lambda_{\rm EDD}$ (bottom panel) in MS galaxies (solid lines). Colors highlight representative cases with different seed masses. For completeness, we also show the extreme case of a continuous SB-like evolution (dashed lines), for which BH and galaxy growth proceed about 5$\times$ faster than on the MS \citep{Schreiber+2015}.

\begin{figure}
     \includegraphics[width=\linewidth]{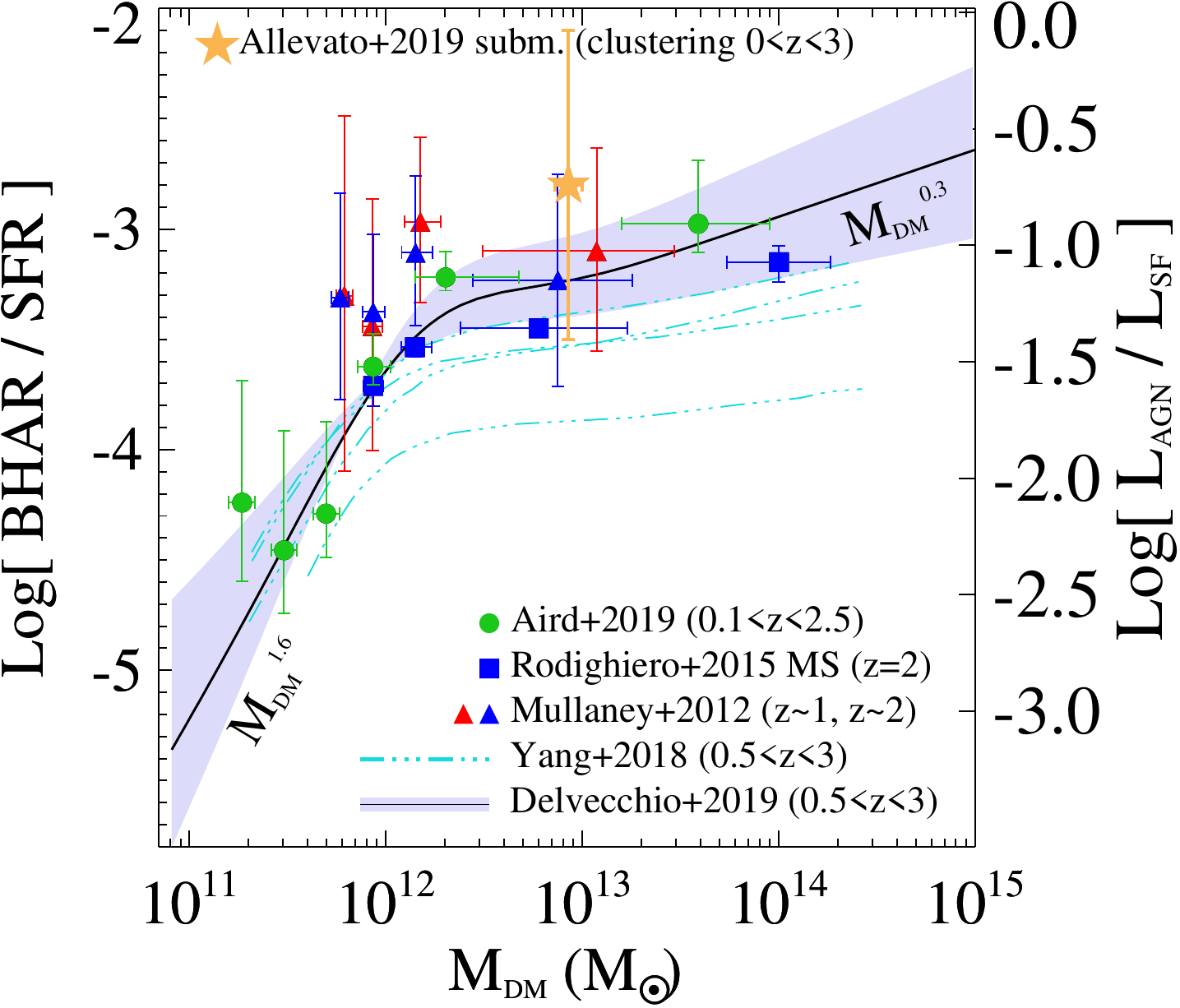}
 \caption{\small Compilation of various BHAR/SFR trends with M$_{\rm DM}$, after applying the M$_{\star}$--M$_{\rm DM}$ conversion for star-forming galaxies \citep{Behroozi+2019} at z=1. Symbols are the same shown in Fig.~\ref{fig:bhsf}. The right y-axis shows the equivalent AGN-to-galaxy bolometric output L$_{\rm AGN}$/L$_{\rm SF}$. The clear twofold trend suggests that M$_{\rm DM}$ might be crucial for explaining the non-linear SMBH growth.
 }
   \label{fig:bhsf_dm}
\end{figure}

The typical M$_{\rm BH}$/M$_{\star}$ ratio of MS galaxies (top panel) decreases from z=10 until z$\sim$2--3, and then rises towards z=0. We note that smaller BH seeds increase their relative mass faster and earlier than bigger BH seeds, at fixed M$_{\star}$. Above a certain critical M$_{\star}$, increasing with seed M$_{\rm BH}$, all seeds converge towards a similar (within a factor of two) M$_{\rm BH}$/M$_{\star}$ ratio at z=0. The evolution of SB galaxies shows instead a minimum at higher redshifts, as they reach the critical M$_{\star}$ on about 5$\times$ shorter timescales ($\propto$M$_{\star}$/SFR) relative to MS analogs. For this reason, the BHAR/SFR ratio of SB galaxies appears systematically higher than for z-matched MS analogs (middle panel). We calculate the mean $\lambda_{\rm EDD}$ from Eq.~\ref{eq:eddratio} and display its evolution with redshift (bottom panel). In MS galaxies, the typical $\lambda_{\rm EDD}$ peaks at 1.5$<$z$<$2.5, slightly increasing with decreasing M$_{\rm BH}$ seed. In comparison, SB galaxies reach a peak at higher redshifts (2$<$z$<$4). 
The blue dotted line indicates the typical fluctuations in $\lambda_{\rm EDD}$ of a SB galaxy with seed (M$_{\rm BH}$,M$_{\star}$) = (10$^{3}$,10$^{7}$)~M$_{\odot}$ (blue dashed line) when propagating the dispersion of the MS relation and the scatter of the assumed BHAR/SFR trend. The amplitude of such fluctuations (similar also for the other mass seeds) suggests that SMBH accretion can vary over several orders of magnitude within the uncertainties, and it may occasionally reach Eddington-limited accretion in starbursting galaxies, consistently with high-redshift model predictions (e.g. \citealt{Lupi+2019}).

The fact that BHAR is enhanced relative to SFR in the most massive galaxies might be also linked to the increasing compactness observed in star-forming galaxies towards higher M$_{\star}$ (M$_{\star}$$\propto$R$^{0.4}$, \citealt{van_der_Wel+2014}). Indeed, a higher compactness might enhance the galaxy ability to retain cold gas re-injected from stellar/AGN feedback, and eventually drive it within the BH sphere of influence.

In addition, environmental mechanisms linked to M$_{\rm DM}$ might help replenish and sustain BH-galaxy growth via inflows of pristine cold gas, predicted to be more effective in massive halos \citep{Dekel+2009}. To test this, we adopt the M$_{\star}$--dependent M$_{\rm DM}$/M$_{\star}$ ratio for star-forming galaxies from \citet{Behroozi+2019} at z=1\footnote{Taking the conversion at a different intermediate redshift would not affect our conclusions.}, and display the BHAR/SFR trend with M$_{\rm DM}$ in Fig.~\ref{fig:bhsf_dm}. The non-linear M$_{\star}$--M$_{\rm DM}$ conversion generates a strikingly twofold behavior that nicely resembles our empirical twofold BH-galaxy growth. In small DM halos Supernovae-driven feedback suppresses BH growth (BHAR/SFR~$\propto$~M$_{\rm DM}^{1.6}$) out to M$_{\rm DM}\sim$2$\times$10$^{12}$~M$_{\odot}$, where baryons are most efficiently converted into stars. Above the turnover M$_{\rm DM}$, AGN activity (exerting both positive and negative feedback) and cold gas inflows might enhance both BH accretion and galaxy star formation. At M$_{\rm DM}\gtrsim$10$^{13}$~M$_{\odot}$ the BHAR/SFR ratio flattens out (BHAR/SFR~$\propto$~M$_{\rm DM}^{0.3}$), possibly due to shock-heated gas within the DM halo \citep{Dekel+2009}. We find a good agreement with the average BHAR/SFR and M$_{\rm DM}$ measurements obtained from clustering of X-ray AGN at 0$<$z$<$3 (Allevato et al. submitted, yellow star). Therefore, we speculate that M$_{\rm DM}$ might be the leading physical driver of the observed non-linear BH-galaxy growth.

In Fig.~\ref{fig:bhsf_dm} we link the BHAR/SFR ratio to the AGN-to-galaxy bolometric output (L$_{\rm AGN}$/L$_{\rm SF}$, right y-axis), assuming that BH accretion occurs with $\epsilon$=0.1. Instead, the galaxy bolometric power arising from star formation (L$_{\rm SF}$) is calculated by converting the (obscured) SFR into rest-frame 8--1000~$\mu$m luminosity, via a \citep{Kennicutt1998} scaling factor. While L$_{\rm AGN}$ is always sub-dominant relative to L$_{\rm SF}$, their ratio increases and displays a turnover in M$_{\rm DM}$ at L$_{\rm AGN}$/L$_{\rm SF}$$\approx$10\%, above which it slowly approaches energy equipartition. Adding some contribution from unobscured star formation, particularly towards low M$_{\star}$, would strenghten the resulting trends. We also note that assuming L$_{\rm X}$--dependent bolometric corrections for deriving the BHAR (e.g. \citealt{Lusso+2012}) would further steepen the resulting BHAR/SFR trend with M$_{\star}$, amplifying the twofold behavior with M$_{\rm DM}$.

We acknowledge that our toy model is not able to reproduce BHs as massive as 10$^{9-10}$~M$_{\odot}$ already at z$\sim$6 (e.g. \citealt{Mortlock+2011}), even in the unlikely scenario of continuous SB-like evolution since z=10. Indeed, our toy model would form BHs with M$_{\rm BH}\lesssim$10$^{8}$~M$_{\odot}$ at z=6, but the galaxy would overgrow in M$_{\star}$ if extrapolating down to z=0 \citep{Renzini2009}. As for local dynamical M$_{\rm BH}$ measurements, we believe that also high-redshift observations are likely biased towards the brightest AGN and most massive BHs that swamp the host-galaxy light, a critical condition to ensures reliable M$_{\rm BH}$ estimates. Therefore, we argue that such quasars at z$\sim$6, if their M$_{\rm BH}$ are not overestimated (but see \citealt{Mejia+2018}), must have grown at a BHAR/SFR ratio about 10$\times$ higher than that assumed in this work. If such a notable ratio was followed by the overall AGN population at z$\sim$6, we would severely overestimate the observed XLF (D19) and the declining BHAR density constrained by deep X-ray data at z$>$3 \citep{Vito+2018}. This leads us believe that the most massive quasars at z$\sim$6 followed very peculiar and uncommon evolutionary paths.

At z=0, Fig.~\ref{fig:z0} shows that our toy model agrees well with proposed de-biased M$_{\rm BH}$--M$_{\star}$ relations \citep{Shankar+2016} and representative local AGN samples \citep{Reines+2015}. Nevertheless, we note a significant ($>$10$\times$) discrepancy at M$_{\star}\lesssim$10$^{11}$~M$_{\odot}$ relative to empirical M$_{\rm BH}$--M$_{\rm bulge}$ relations based on dynamical M$_{\rm BH}$ measurements (\citealt{Kormendy+2013}). This apparent conflict might arise from multiple reasons: (i) The local M$_{\rm BH}$--M$_{\rm bulge}$ relation is likely biased towards the largest BHs hosted within massive quiescent systems, for which the BH sphere of influence can be spatially resolved (\citealt{Gultekin+2009}; \citealt{Shankar+2016}, \citeyear{Shankar+2019}). This biases the intrinsic M$_{\rm BH}$--M$_{\rm bulge}$ relation towards a flatter slope and higher normalization \citep{Volonteri+2011}. (ii) While in the most massive galaxies M$_{\rm bulge}\approx$~M$_{\star}$, at M$_{\star}$=10$^{10}$~M$_{\odot}$ the B/T decreases to $\approx$0.3 for local MS galaxies \citep{Dimauro+2018}. This behavior causes a M$_{\star}$--dependent steepening of the M$_{\rm BH}$--M$_{\rm bulge}$ relation (green dashed line in Fig.~\ref{fig:z0}), though not crucial for reconciling the observed difference. (iii) A significant fraction of SMBH accretion might be heavily obscured, thus unaccessible via X-ray observations. However, this elusive contribution might boost the average BHAR by at most factor of $\lesssim$2, thus not filling the observed gap at low M$_{\star}$ \citep{Comastri+2015}. (iv) The average radiative efficiency might be much lower than $\epsilon$=0.1 (possibly M$_{\star}$ dependent). As a consequence, the corresponding mass accreted by SMBHs at fixed luminosity would be higher. However, the lowest theoretically-expected value of $\epsilon$=0.06 \citep{Novikov+1973} proves still insufficient to justify the observed discrepancy. Therefore, we favor the combination of points (i) and (ii) as possible reasons to explain the conflicting M$_{\rm BH}$--M$_{\star}$ trends at z=0.

Concluding, the proposed empirically-motivated BHAR/SFR trend with M$_{\star}$ (D19) enables us to describe the cosmic SMBH-galaxy assembly in normal SF galaxies, in agreement with high-z cosmological simulations and intrinsic M$_{\rm BH}$--M$_{\star}$ relations at z=0. Our study suggests that the DM halo mass primarily regulates the amount of cold gas available for triggering and sustaining the cosmic non-linear BH-galaxy growth.

\bigskip
\acknowledgments{We thank the anonymous referee for a quick and constructive report. We are grateful to Paola Dimauro for sharing the median bulge-to-total ratio of MS galaxies, and Viola Allevato for sharing the clustering-based BHAR/SFR. ID is supported by the European Union's Horizon 2020 research and innovation program under the Marie Sk\l{}odowska-Curie grant agreement No~788679. FS acknowledges partial support from a Leverhulme Trust Fellowship. AC acknowledges the support from grants PRIN MIUR 2015 and 2017. }

\bibliographystyle{aasjournal}

\begin{thebibliography}{}
 \expandafter\ifx\csname natexlab\endcsname\relax\def\natexlab#1{#1}\fi

\bibitem[Aird et al.(2012)]{Aird+2012} Aird, J., Coil, A.~L., Moustakas, J., et al.\ 2012, \apj, 746, 90

\bibitem[Aird et al.(2015)]{Aird+2015} Aird, J., Alexander, D.~M., Ballantyne, D.~R., et al.\ 2015, \apj, 815, 66 

\bibitem[Aird et al.(2019)]{Aird+2019} Aird, J., Coil, A.~L., \& Georgakakis, A.\ 2019, \mnras, 484, 4360 

\bibitem[Ba{\~n}ados et al.(2018)]{Banados+2018} Ba{\~n}ados, E., Venemans, B.~P., Mazzucchelli, C., et al.\ 2018, \nat, 553, 473 

\bibitem[Begelman, \& Rees(1978)]{Begelman+1978} Begelman, M.~C., \& Rees, M.~J.\ 1978, \mnras, 185, 847

\bibitem[Behroozi et al.(2013)]{Behroozi+2013} Behroozi, P.~S., Wechsler, R.~H., \& Conroy, C.\ 2013, \apj, 770, 57

\bibitem[Behroozi et al.(2019)]{Behroozi+2019} Behroozi, P., Wechsler, R.~H., Hearin, A.~P., et al.\ 2019, \mnras, 488, 3143

\bibitem[Bower et al.(2017)]{Bower+2017} Bower, R.~G., Schaye, J., Frenk, C.~S., et al.\ 2017, \mnras, 465, 32 

\bibitem[Chabrier(2003)]{Chabrier2003} Chabrier, G.\ 2003, \apjl, 586, L133 

\bibitem[Comastri et al.(2015)]{Comastri+2015} Comastri, A., Gilli, R., Marconi, A., Risaliti, G., \& Salvati, M.\ 2015, \aap, 574, L10 

\bibitem[Dekel et al.(2009)]{Dekel+2009} Dekel, A., Birnboim, Y., Engel, G., et al.\ 2009, \nat, 457, 451 

\bibitem[Delvecchio et al.(2019)]{Delvecchio+2019} Delvecchio et al., to be submitted, 2019, \apj 

\bibitem[Dimauro et al.(2018)]{Dimauro+2018} Dimauro, P., Huertas-Company, M., Daddi, E., et al.\ 2018, \mnras, 478, 5410 

\bibitem[Dubois et al.(2014)]{Dubois+2014} Dubois, Y., Volonteri, M., \& Silk, J.\ 2014, \mnras, 440, 1590 

\bibitem[Ferrarese \& Merritt(2000)]{Ferrarese+2000} Ferrarese, L., \& Merritt, D.\ 2000, \apjl, 539, L9 

\bibitem[Gebhardt et al.(2000)]{Gebhardt+2000} Gebhardt, K., Bender, R., Bower, G., et al.\ 2000, \apjl, 539, L13 

\bibitem[G{\"u}ltekin et al.(2009)]{Gultekin+2009} G{\"u}ltekin, K., Richstone, D.~O., Gebhardt, K., et al.\ 2009, \apj, 698, 198 

\bibitem[Habouzit et al.(2017)]{Habouzit+2017} Habouzit, M., Volonteri, M., \& Dubois, Y.\ 2017, \mnras, 468, 3935 

\bibitem[H{\"a}ring \& Rix(2004)]{Haring+2004} H{\"a}ring, N., \& Rix, H.-W.\ 2004, \apjl, 604, L89 

\bibitem[Kennicutt(1998)]{Kennicutt1998} Kennicutt, R.~C., Jr.\ 1998, \apj, 498, 541 

\bibitem[Kormendy \& Ho(2013)]{Kormendy+2013} Kormendy, J., \& Ho, L.~C.\ 2013, \araa, 51, 511 

\bibitem[Leitner, \& Kravtsov(2011)]{Leitner+2011} Leitner, S.~N., \& Kravtsov, A.~V.\ 2011, \apj, 734, 48

\bibitem[Lupi et al.(2019)]{Lupi+2019} Lupi, A., Volonteri, M., Decarli, R., et al.\ 2019, \mnras, 488, 4004 

\bibitem[Lusso et al.(2012)]{Lusso+2012} Lusso, E., Comastri, A., Simmons, B.~D., et al.\ 2012, \mnras, 425, 623 

\bibitem[Magorrian et al.(1998)]{Magorrian1998} Magorrian, J., Tremaine, S., Richstone, D., et al.\ 1998, \aj, 115, 2285 

\bibitem[Marconi et al.(2004)]{Marconi+2004} Marconi, A., Risaliti, G., Gilli, R., et al.\ 2004, \mnras, 351, 169 

\bibitem[McAlpine et al.(2018)]{McAlpine+2018} McAlpine, S., Bower, R.~G., Rosario, D.~J., et al.\ 2018, \mnras, 481, 3118 

\bibitem[Mej{\'\i}a-Restrepo et al.(2018)]{Mejia+2018} Mej{\'\i}a-Restrepo, J.~E., Lira, P., Netzer, H., et al.\ 2018, Nature Astronomy, 2, 63

\bibitem[Merloni et al.(2010)]{Merloni+2010} Merloni, A., Bongiorno, A., Bolzonella, M., et al.\ 2010, \apj, 708, 137 

\bibitem[Mortlock et al.(2011)]{Mortlock+2011} Mortlock, D.~J., Warren, S.~J., Venemans, B.~P., et al.\ 2011, \nat, 474, 616 

\bibitem[Mullaney et al.(2012)]{Mullaney+2012} Mullaney, J.~R., Daddi, E., B{\'e}thermin, M., et al.\ 2012, \apjl, 753, L30 

\bibitem[Netzer et al.(2014)]{Netzer+2014} Netzer, H., Mor, R., Trakhtenbrot, B., Shemmer, O., \& Lira, P.\ 2014, \apj, 791, 34 

\bibitem[Novikov \& Thorne(1973)]{Novikov+1973} Novikov, I.~D., \& Thorne, K.~S.\ 1973, Black Holes (Les Astres Occlus), 343 

\bibitem[Reines \& Volonteri(2015)]{Reines+2015} Reines, A.~E., \& Volonteri, M.\ 2015, \apj, 813, 82 

\bibitem[Renzini(2009)]{Renzini2009} Renzini, A.\ 2009, \mnras, 398, L58 

\bibitem[Rodighiero et al.(2011)]{Rodighiero+2011} Rodighiero, G., Daddi, E., Baronchelli, I., et al.\ 2011, \apjl, 739, L40 

\bibitem[Rodighiero et al.(2015)]{Rodighiero+2015} Rodighiero, G., Brusa, M., Daddi, E., et al.\ 2015, \apjl, 800, L10 

\bibitem[Soltan(1982)]{Soltan1982} Soltan, A.\ 1982, \mnras, 200, 115 

\bibitem[Schaye et al.(2015)]{Schaye+2015} Schaye, J., Crain, R.~A., Bower, R.~G., et al.\ 2015, \mnras, 446, 521

\bibitem[Schreiber et al.(2015)]{Schreiber+2015} Schreiber, C., Pannella, M., Elbaz, D., et al.\ 2015, \aap, 575, A74 

\bibitem[Shankar et al.(2016)]{Shankar+2016} Shankar, F., Bernardi, M., Sheth, R.~K., et al.\ 2016, \mnras, 460, 3119 

\bibitem[Shankar et al.(2019)]{Shankar+2019} Shankar, F., Bernardi, M., Richardson, K., et al.\ 2019, \mnras, 485, 1278 

\bibitem[van der Wel et al.(2014)]{van_der_Wel+2014} van der Wel, A., Franx, M., van Dokkum, P.~G., et al.\ 2014, \apj, 788, 28 

\bibitem[Vito et al.(2018)]{Vito+2018} Vito, F., Brandt, W.~N., Yang, G., et al.\ 2018, \mnras, 473, 2378 

\bibitem[Vito et al.(2019)]{Vito+2019} Vito, F., Brandt, W.~N., Bauer, F.~E., et al.\ 2019, \aap, 628, L6 

\bibitem[Volonteri(2010)]{Volonteri2010} Volonteri, M.\ 2010, \aapr, 18, 279 

\bibitem[Volonteri \& Stark(2011)]{Volonteri+2011} Volonteri, M., \& Stark, D.~P.\ 2011, \mnras, 417, 2085 

\bibitem[Wang et al.(2015)]{Wang+2015} Wang, F., Wu, X.-B., Fan, X., et al.\ 2015, \apjl, 807, L9 

\bibitem[Wu et al.(2015)]{Wu+2015} Wu, X.-B., Wang, F., Fan, X., et al.\ 2015, \nat, 518, 512 

\bibitem[Yang et al.(2018)]{Yang+2018} Yang, G., Brandt, W.~N., Vito, F., et al.\ 2018, \mnras, 475, 1887 



 \end{thebibliography}

\end{document}